\documentclass{nlsart}
\usepackage{graphicx}
\begin{document}
\begin{frontmatter}

\title{Muon $g-2$ and New Physics}

\thanks[talk]
       {Talk given at Novosivirsk,
        ``International Workshop $e^+e^-$ Collisions
        from $\phi$ to $J/\psi$'', March 1 - 5, 1999
        (Novosivirsk, March 1999)}

\author[Hayakawa]{M. Hayakawa}
\thanks[Hayakawa]{electric address: haya@mbox.post.kek.jp}

\address{Theory Division, KEK,
         Tsukuba, Ibaraki 305-0801, Japan \\
         Department of Physics, Tokyo Institute of Technology, \\
         \={O}okayama, Tokyo 152-0033, Japan}

\begin{abstract}
  Here we invoke
the current and future perspective on muon $g-2$ measurement
when asking what
the muon $g-2$ could tell about the underlying structure concerning
with the hierarchy problem.
  Here we take up two such models,
the presence of which turns out to alter
the standard model prediction for muon $g-2$ significantly:
one is TeV scale gravity scenario,
the other supersymmetric model,
in the latter case of which
the precision measurement up to $Z$ boson mass is taken into account
as an explicit constraint.
\end{abstract}

\end{frontmatter}

\section{Introduction}
\label{sec:Intro}
\quad
 Further precise measurement of
the anomalous magnetic dipole moment of the muon,
conventionally denoted as $a_\mu \equiv \frac{1}{2}(g-2)_\mu$,
is now underway
at Brookhaven National Laboratory (BNL).
 The perspective for the goal of this experiment is
\cite{future-exp}
\begin{equation}
  \Delta a_\mu({\rm expt}) =
  4.0 \times 10^{-10}\, .
   \label{eq:future-precision}
\end{equation}
 The recent report for its test-running
course at BNL \cite{future-exp,BNL-NV}
combined with the previous one at CERN \cite{CERN}
gives to muon $g-2$
\begin{equation}
  a_\mu(\textrm{expt}) = 11659~235~(73) \times 10^{-10}\, ,
   \label{eq:present_value}
\end{equation}
where the numerals in the parenthesis represent
the uncertainty in the final few digits.
 Thus the precision (\ref{eq:future-precision})
amounts to the determination of its value by one further digit.
\\
\quad
 Our primary interest
is what we can learn when invoking
such a improvement in muon $g-2$ experiment.
 At present the standard model predicts
\begin{equation}
  a_\mu(\textrm{SM}) =
   11659~160.5~(6.5) \times 10^{-10} ,
\end{equation}
which includes the up-dated estimate on
the leading hadronic vacuum polarization contribution
\cite{Davier},
(See \cite{Eidelman} due to analysis without recourse to $\tau$ decay data),
and $\mathcal{O}(\alpha^3)$ contribution
\cite{Krause}.
  The designed precision (\ref{eq:future-precision})
is put forward to find out the existence
of the $W$, $Z$ boson effect to muon $g-2$
\cite{weak-2-loop}
\begin{equation}
  a_\mu(\textrm{weak}) =
   151~(4) \times 10^{-11} .
  \label{eq:weak_effect}
\end{equation}
  The electron $g-2$ does not receive observable effects
from $Z$ and $W$ bosons
and are entirely saturated by QED effect.
  This fact enables us to find out the validity of
calculation scheme of quantum field theory,
including the perturbative renormalization procedure.
  Rather electron $g-2$ provides
the most accurate determination of the fine structure constant
$\alpha$ \cite{Kinoshita} at present.
  From (\ref{eq:weak_effect}) and (\ref{eq:future-precision}),
the muon $g-2$ is expected not only to obtain
a conceivable evidence for
such a structure about the electroweak interactions
as involved in standard model,
but also has the potential
to indicate the existence of much richer ingredient
associated with some theoretical problem. \\
\quad
 When we incline to use muon $g-2$
as a probe of new physics,
the theme of the talk assigned to me as well,
it would be important to recall
the motivation or the merits of each model.
 Thus, here, I will focus on two concrete models
considered to approach to ''hierarchy problem'',
although it is somewhat a conceptual viewpoint.
 Here ''hierarchy problem'' stands for
the following question in some narrow sense;
what is assuring the stability of the electroweak scale,
represented by $W$ boson mass, $M_W$,
against quantum fluctuation associated with
high momentum modes below some cutoff scale.
 The cutoff scale here is the scale
at which the gauge interactions appearing in standard model would
become subject to some kind of modification.
 It may be the GUT scale $M_G$,
at which the standard model gauge symmetry group
is merged into a larger symmetry group.
 Or it may be the scale at which the gauge boson is resolved
into more fundamental structure, for instance, into string. \\
\quad
 Here we take up two models considered with such a motivation.
 One is TeV scale gravity discussed in the next section,
and the other supersymmetry in Sec. \ref{sec:SUSY}.
 Sec. \ref{sec:discussion} concludes
with remark on several facets
for muon $g-2$ to probe new physics practically.
\section{TeV scale gravity}
\label{sec:gravity}
\quad
 The laboratory experiments check the structure of
gravity has been met up to the order of millimeters.
 With this in mind let us turn our attention
to the scheme introduced a couple of years ago
\cite{Dimopoulos}. \\
\quad
 Let us imagine that our world is confined in a three-brane,
the extended object with three-spatial directions,
flowing in a higher dimensional space
($(n+4)$-dimension in total).
 These $n$-directions are compactified to an $n$-dimensional torus
with the same length $2\pi R$, for simplicity.
 Then there are an infinite tower of Kaluza-Klein states
from the four-dimensional space-time point of view.
 They are the states with non-zero momentum in the extra directions.
 In four-dimensional world
such a momentum appears as the mass,
the scale of which is characterized by inverse of $R$. \\
\quad
 The behavior of static potential between two point-like sources
prepared with separation $r$
illustrates an important aspect of the present setting
\cite{Arkani-Hamed} .
 When $r$ is large compared to $R$,
the potential behaves as $\sim 1/r$,
nothing but the form of the usual Newtonian one.
 The Planck scale $M_{\rm PL} \simeq 10^{19}$ characterizes
its strength.
 However, once the separation $r$ reaches below $R$,
the probability that Kaluza-Klein states get excited
cannot become neglected any more.
 Thus counting those states
which propagate essentially like massless states
between two sources
leads $r^{-(n+1)}$ as
$r$-dependence of the potential at short distance,
characteristic of $(n+3)$-spatial dimension.
 This only reflects the fact
that the local structure less than the compactification scale
is that of $(n+4)$-dimensional space-time.
 A noteworthy point is
that the strength of force is
then characterized by another Planck scale,
$M_*$;
\begin{equation}
 M_* =
  \left(
    \frac{M_{\rm PL}^2}{R^n}
  \right)^{(n+2)}\, .
\end{equation}
 This is
the strength of the gravitational interaction in the bulk theory,
the fundamental Planck scale.
\\
\quad
 Now we take $R$ equal to 1 mm
\footnote{
 Energy scale is related to the length scale $L$ through
\begin{equation}
 E = 0.197\,{\rm eV} \times \frac{1\,{\rm mm}}{L}\, .
  \nonumber
\end{equation} 
},
which corresponds to the length scale one-order less than
the current reach of the experiment on gravity.
 With two extra compactified directions
$M_*$ = 1 TeV
\footnote{
 It was commented by A. Kataev in this workshop
that consideration on the effect
to the life-time of the red giant stars
appears to reject the possibility, $M_*$ = 1 TeV,
which he heard at the seminar by Arkani-Hamed
\cite{Arkani-Hamed}.}.
\\
\quad
 If the three-brane were further of Dirichlet-brane type,
on which the open string can attach,
our standard model gauge bosons
would become the tangential component of
the ground state of such an open string.
 Then the gauge bosons
would get resolve into strings higher than 1 TeV
as long as the string coupling constant is of order unity.
 Thus there is no hierarchy problem ab initio.
\\
\quad
 At present we are lacking the precise
formulation of theories and
the detail knowledge on its dynamical aspects
(especially on the compactification mechanism).
 However the dimension counting argument
with symmetry consideration
has been an enough tool
when we estimate the order of magnitude
about some effect in low energy phenomena
unless the effective theory description breaks down,
although we rather expect to superstring theory
that many miracles beyond this assumption occur.
\\
\quad
 The argument begins with the chiral symmetry for muon.
 When muon was massless, the magnetic dipole coupling would be absent.
 Thus the magnitude of magnetic moment will
be proportional to muon mass $m_\mu$
and the corresponding operator appears
in the effective Lagrangian:
\begin{equation}
 \mathcal{L} = e\,m_\mu \,A\,\bar{\mu}\,
               \sigma^{\lambda\rho} F_{\lambda\rho}\, \mu.
  \label{eq:muon_dipole}
\end{equation}
with the the extra effect $A$ due to the new structure
characterized by mass scale $M_*$.
 Since the mass dimension of $A$ turns out to be minus one
in four space-time dimension,
the dimension counting now gives
\begin{equation}
 A = c \times \frac{1}{M_*},
  \label{eq:A_const}
\end{equation}
with the numerical constant $c$ of order unity.
 Insertion of (\ref{eq:A_const}) into
the effective interaction (\ref{eq:muon_dipole})
with a slight rearrangement yields
\begin{equation}
 \mathcal{L} =
  \frac{e}{4 m_\mu}\,
  \left[
   4c\,
   \left(
    \frac{m_\mu}{M_*}
   \right)^2
  \right]\,
  \bar{\mu} i\sigma^{\lambda\rho} F_{\lambda\rho}\,\mu.
\end{equation}
 The quantity found in the square bracket of the above expression
corresponds to the additional contribution
to $a_\mu$ due to the presence of TeV scale gravity.
 Thus the additional effect to $a_\mu$
can be read off as
\begin{equation}
 \delta a_\mu
  = 4c\,
    \left( \frac{m_\mu}{M_*} \right)^2
    \times 10^{10},
\end{equation}
which becomes for $M_*$ = 1 TeV
\begin{equation}
 \delta a_\mu \sim
  \left( 4c \times 100 \right) \times 10^{-10}. 
\end{equation}
 Thus a crude estimate shows that
the effect from TeV scale gravity is in the marginal
situation to be detected even
with the current accuracy (\ref{eq:present_value}).
 The future accuracy (\ref{eq:future-precision})
is quite adequate to detect the existence
of new aspect of gravity characterized by TeV scale
\footnote{
 During the preparation of the talk,
I have noticed that more concrete demonstration
has been performed in a similar context
\cite{Graesser}
}.
\section{Supersymmetric Model}
\label{sec:SUSY}
\begin{figure}[thb]
 \begin{center}
  \includegraphics[origin=Bc, scale=0.9]{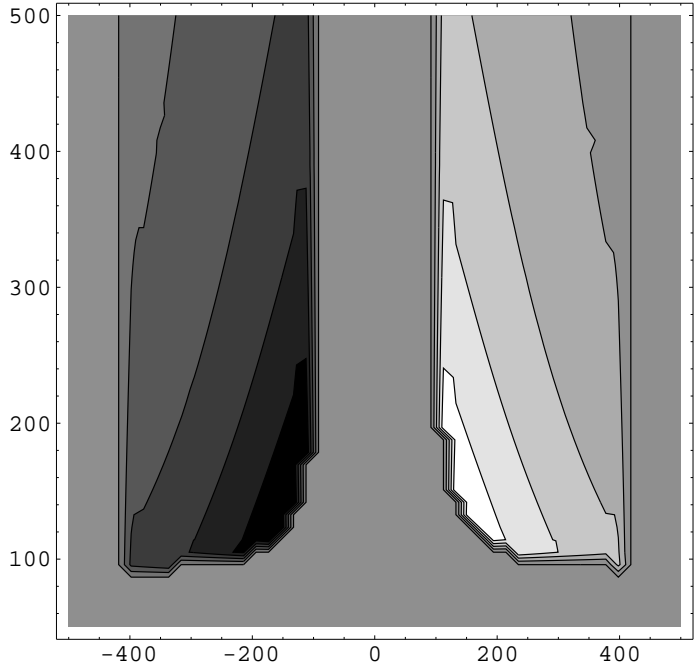}
  \caption{Contours with equal $(a_\mu)_{\rm SUSY}$
          in $\mu$(horizontal direction)-$M_2$ plane
          for $\tan\beta = 50$ and
          $m_{\bar{\mu}_L} = m_{\bar{\mu}_R} = 200$ GeV.
          The two islands are the regions escaping from
          any other constraints.
          The contours are drawn with the interval of
          $50 \times 10^{-10}$
          from $-200 \times 10^{-10}$ to $200 \times 10^{-10}$
          for $(a_\mu)_{\rm SUSY}$.
          Darker face corresponds to
          smaller $(a_\mu)_{\rm SUSY}$.}
  \label{fig:cnt_50_200}
 \end{center}
\end{figure}

\begin{figure}[htb]
 \includegraphics[origin=Bc, angle=-90, scale=0.6]{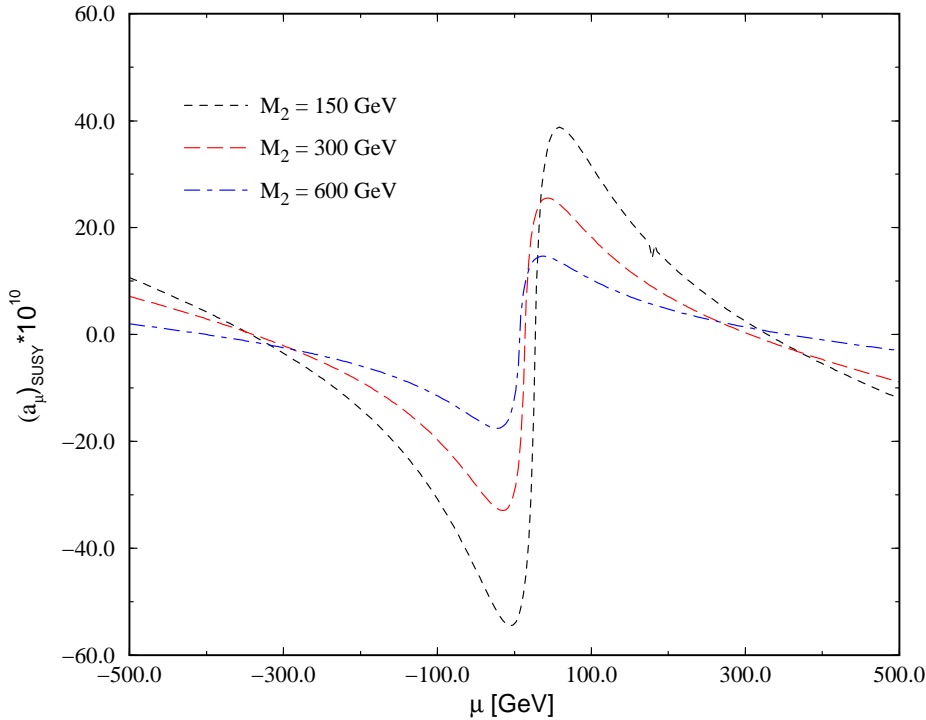}
 \caption{$\mu$ dependence of $(a_\mu)_{\rm SUSY}$
          for $\tan\beta = 3$ and
          $m_{\tilde{\mu}_L} = m_{\tilde{\mu}_R} = 100$ GeV.}
 \label{fig:g-2_3_100}
\end{figure}

\begin{figure}[htb]
 \begin{center}
  \includegraphics[origin=Bc, scale=0.9]{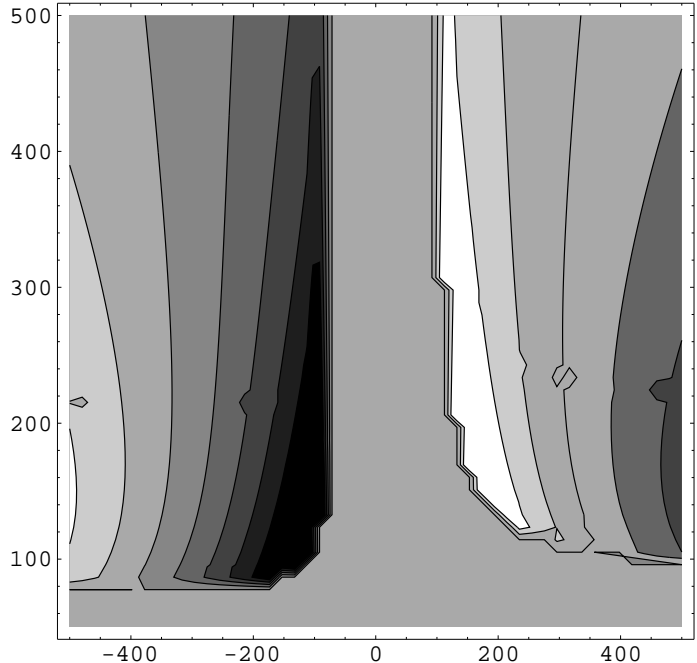}
  \caption{Similar contours with equal $(a_\mu)_{\rm SUSY}$
          in $\mu$-$M_2$ plane
          for $\tan\beta = 3$ and
          $m_{\bar{\mu}_L} = m_{\bar{\mu}_R} = 100$ GeV.          
          The contours are drawn with the interval of
          $5\times 10^{-10}$
          between $-25 \times 10^{-10}$ and $10 \times 10^{-10}$
          for $(a_\mu)_{\rm SUSY}$.}
 \label{fig:cnt_3_100}
 \end{center}
\end{figure}

\quad
 Now we will take our attention to the search of supersymmetry
with the use of muon $g-2$,
which has been discussed
as a machinary to check the various adovocated models
or from some generic standpoint
\cite{Kosower,Chattopadhyay,Polonsky,Gabrielli}.
 The detail formula and so on in this section
will also be found in a separate literature
\cite{CHH}.
\\
\quad
 In the narrow sense
of the ''hierarchy problem'' defined in the introductory remark,
supersymmetry makes it possible
to extend the standard model gauge group into
the larger gauge group in grand unified theory
(GUT).
 This is realized in the form of
the cancellation of the dangerous quantum corrections
within each supermultiplet.
 The bosonic partner of muon,
for instance, has not been observed yet
so that it must be much heavier.
 Smuon can be let heavier by giving it a lifting-up mass, $m_S$.
 This procedure does not spoil quantum stability
as far as $m_S$ is not so apart from the electroweak scale.
 The scale $m_S$ works
as the ultraviolet cutoff scale for the phenomena
accessed by low momentum probe
while it works as the infrared cutoff
from the supersymmetric high-energy side.
 It has been recognized that
unification of the coupling constants of three gauge interactions,
$\textrm{SU(2)}_{\rm L}$,  $\textrm{U(1)}_{\rm Y}$
and color $\textrm{SU(3)}_{\rm C}$,
is achieved by assuming the particle content
of the minimal supersymmetric extension of standard model
above such $m_S$.
 For the future purposes it deserves to recall here
that the knowledge from precise
measurement around $Z$ pole plays an indispensable role
to establish this fact.
\\
\quad
 Now we examine the effect on muon $g-2$
induced from supersymmetric theories.
 They come essentially from two diagrams.
 One is the chargino-sneutrino loop,
where the charginos are the admixture of $\textrm{SU(2)}$
gaugino $\tilde{w}^-$ and the charged Higgsino.
 The other is the neutralino-smuon loop,
where the neutralinos are the admixture of
their neutral counterparts.
 The other contributions,
such as the charged Higgs loop one,
are so small that they are irrelevant
even for future study.
\\
\quad
\begin{figure}[htb]
 \includegraphics[origin=Bc, angle=-90, scale=0.6]{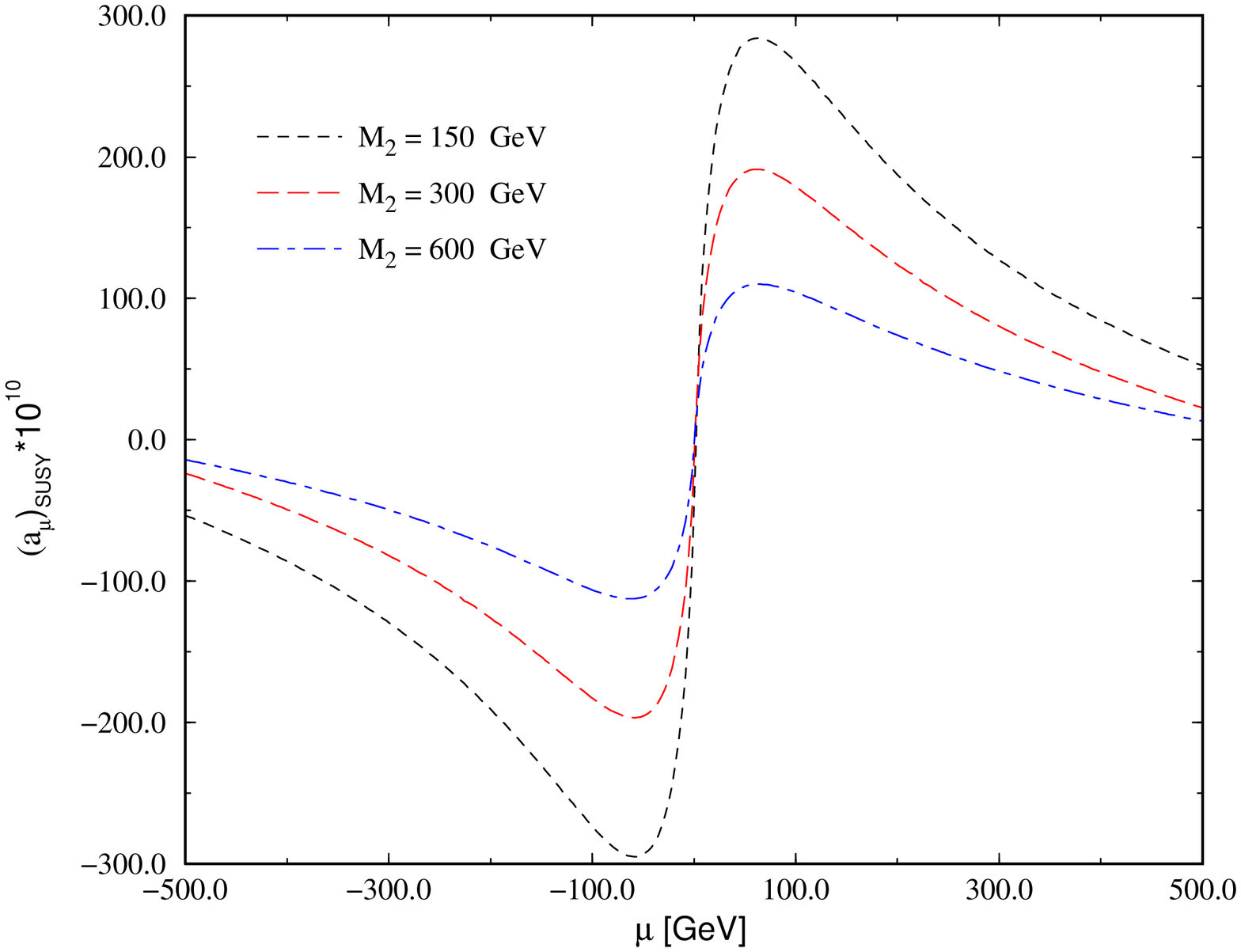}
 \caption{$\mu$ dependence of $(a_\mu)_{\rm SUSY}$
          for $\tan\beta = 50$ and
          $m_{\tilde{\mu}_L} = m_{\tilde{\mu}_R} = 200$ GeV.}
 \label{fig:g-2_50_200}
\end{figure}
 Fig. \ref{fig:g-2_50_200}
is intended to demonstrate
what magnitude of those effects to muon $g-2$
is expected.
 The figure shows
the supersymmetric contribution to muon $g-2$
as a function of $\mu$ parameter
(the supersymmetric mass common to two Higgs supermultiplets),
for relatively large $\tan\beta$
(the ratio of the vacuum expectation values of two Higgs doublets),
the supersymmetry breaking slepton mass set equal to 200 GeV,
and for three choices of supersymmetry breaking ${\rm SU(2)_L}$
gaugino mass $M_2$ with ${\rm U(1)_Y}$ gaugino mass given
here through the GUT relation
\begin{equation}
 M_1 = \frac{5}{3} \tan^2\theta_W M_2\, .
\end{equation}
 As the weak effect is $15 \times 10^{-10}$,
the supersymmetric effect can become substantial.
 Actually the muon $g-2$ even with the current accuracy
excludes the region of negative sign of $\mu$
for this set of the other parameters.
\\
\quad
 Fig. \ref{fig:g-2_50_200} is drawn without paying any attention
to the other constraints on supersymmetric models already present.
 The direct search of superpartners of the known species
puts the lowest bound (93 GeV) to the lightest chargino mass,
and the bound (78 GeV) to the mass of each lighter slepton
\cite{Mihara}.
 In fact the chargino mass bound requires that
the absolute magnitude of $\mu$ parameter be greater than
about 100 GeV
for the gaugino mass in our interest,
while the slepton mass bound demands $\left| \mu \right|$
less than about 400 GeV.
 With those regions excluded
Fig. \ref{fig:cnt_50_200} shows
the contours each of which has equal magnitude of muon $g-2$
on the $\mu$-$M_2$ plane.
 Since the future accuracy is much smaller than
the interval between the neighboring contours
in Fig. \ref{fig:cnt_50_200},
we have a great chance to observe a signal
coming from the existence of supersymmetry through muon $g-2$.
\\
\quad
 Now we are tempted to grasp the specific feature of muon $g-2$
in search of supersymmetric theory.
 It will turn out that muon $g-2$ seems to have a peculiar property
which is not shared by any other observables.
\\
\quad
 In the most honest region of parameter space
$\left| \mu \right| \ge (2 \sim 3)\, M_2$,
chargino-sneutrino loop contribution dominates over
neutralino-smuon loop contribution.
 The current basis analysis
helps one to catch up with the qualitative dependence on
the various parameters.
 As long as $\tan\beta \ge 3$,
the chirality flips due to the vacuum expectation value
$ \left< H_U \right> $
of the Higgs field giving mass to the up-type quarks,
turning $\tilde{w}^-$ to the charged component of $\tilde{H}_U$,
which transformed to the charged $\tilde{H}_D$ due to $\mu$-term.
 Picking $\sin\beta$ from $ \left< H_U \right> $
and $1/\cos\beta$ from a yukawa-type coupling involving muon,
the dominant contribution in the present situation
becomes
\begin{equation}
 (a_\mu)_{\rm SUSY} \propto +\mu \tan\beta \, ,
  \label{eq:a_susy_1}
\end{equation}
although the overall sign needs a detail computation.
 From this expression we can read off such properties that
\begin{enumerate}
 \item[(a)]
  The effect to muon $g-2$
 is greatly enhanced for large $\tan\beta$ \cite{Kosower}.
  In fact, when $\tan\beta$ is small,
 the overall magnitude of SUSY effect is drastically
 reduced as shown in Fig. \ref{fig:g-2_3_100}.
  Thus in this case the current experiment
 could not put any restriction on its existence.
  But the future accuracy in muon $g-2$
 is quite sufficient to explore it
 \footnote{
  The renomalization group analysis shows that
 QED correction tends to decrease those new effect
 about 6\%, and this fact should be recalled
 at the critical stage of confronting
 with the experimental data \cite{Degrassi}.
 }.
 \item[(b)]
  The sign of this contribution is govern by the sign of $\mu$.
\end{enumerate}
 It is interesting to remind that
a large $\tan\beta$ is a natural consequence
of the gauge mediated SUSY breaking scenario,
and elaborate analysis on muon $g-2$
has been performed in this line \cite{Gabrielli}.
 Or it is a necessary ingredient
for unification of all yukawa coupling constants
of the third generation.
\\
\quad
 As observed at both ends of $\mu$-direction
in Fig. \ref{fig:g-2_3_100},
supersymmetric effect to muon $g-2$ does not decouple
even if we can let the absolute magnitude of $\mu$ large
while those other parameters remain fixed.
 Note that small $\tan\beta$ case allows
relatively large absolute magnitude of $\mu$ parameter
without conflicting with slepton mass bound,
as the mixing between left- and right-sleptons are
proportional to a combination $\mu \tan\beta$.
\\
\quad
 Such a phenomenon can be understood from the following observation.
 When $\left| \mu \right|$ is large
the chargino-sneutrino effect decouples,
but the neutralino-smuon effect increases.
 Let us consider a diagram in which
the chiral flip occurs due to the mixing between the left
and right-handed smuons in the current eigen-basis.
 As the Higgino does not propagate,
suppression factor due to the inverse power of $\mu$ is now absent. 
 Thus $(a_\mu)_{\rm SUSY}$ becomes proportional to $-\mu\tan\beta$.
(The sign is also opposite to the chargino-sneutrino effect
(\ref{eq:a_susy_1}).)
 This is the reason for such a behavior
in this large $\left| \mu \right|$ region.
\\
\quad
 From those observations,
muon $g-2$ seems to play the major role
to find the sign and the magnitude of $\mu$ term.
 As far as I know, such a property sensible to $\mu$
is not shared by any other observables.
 Recall the following two facts, that is,
\begin{enumerate}
  \item[(a)] $\mu$ is a supersymmetric parameter
   which is associated with common mass of Higgs supermultiplet.
    Thus this is a parameter independent of
   the supersymmetry breaking parameters by nature.
  \item[(b)] Although supersymmetry
   assures the quantum stability of electroweak scale,
   supersymmetry does not set $\mu$ to this region
   at tree level automatically.
    This is the most annoying matter called as ``$\mu$ problem.
   This problem stands out especially in the context of GUT.
\end{enumerate}
 Thus, once supersymmetry is established
also by the other experiments,
the determination of $\mu$ parameter through muon $g-2$
may develop further theoretical access
to the origin of $\mu$, the origin of electroweak scale.
\\
\quad
 Before addressing to the future testing possibility
in the small $\tan\beta$ regime,
we remind the additional constraints
implied from precision measurement at $Z$ pole.
 As was mentioned,
the result of this precision measurement
has given an indispensable information
to argue grand unification.
 It also has killed the naive technicolor models.
 Thus we should discuss the effect on muon $g-2$
on the region of the parameter space 
consistent with those measurements.
\\
\quad
 They are summarized by four parameters.
 Three of these parametrizes the ``oblique'' corrections
from new physics,
with respect to ``reference'' standard model;
here we take the one specified by
\begin{equation}
 m_t = 175\ {\rm GeV}, \quad
 M_H = 100\ {\rm GeV}\, .
\end{equation}
 The last one is associated
with the modification of coupling of bottom quark to $Z$ boson.
 This is neglected here
by assuming that the squarks are so heavy enough
that their effects decouple.
 Since it has been recognized that
the SUSY effect to $W$ boson mass and coupling of $\tau$ to $Z$
is not relevant within the current accuracy\cite{Cho},
we concentrate on $S$ and $T$ parameters.
\begin{figure}[htb]
 \begin{center}
  \includegraphics[origin=Bc, scale=0.6]{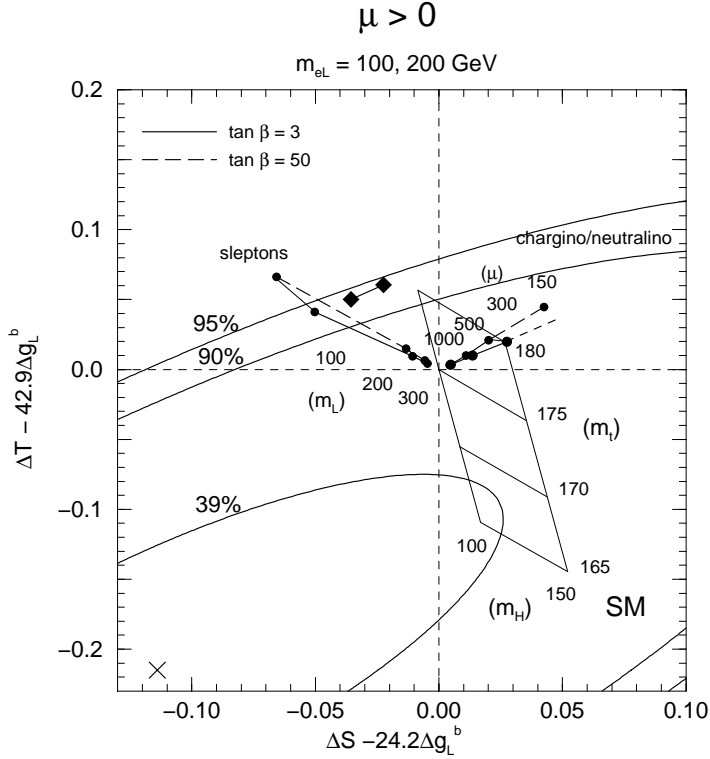}
  \caption{Constraint on supersymmetric theory
          from $S$-$T$ parameters for $\mu > 0$.
          Slepton brings $S$ parameter to decrease
          (Each line follows the response to the change
          of $m_{\tilde{l}_L}$.)
          while chargino and neutralino tend to
          increase it
          (Each line shows the response to the change of $\mu$
          parameter.).
          Therefore they
          partly cancels when added together
          (the line with square dots
          for $m_{\tilde{l}_L}$ = 100 GeV,
          with triangle dots for $m_{\tilde{l}_L}$ = 200 GeV
          for $\tan\beta$ = 3).}
   \end{center}
 \label{fig:ST_100}
\end{figure}
\\
\quad
 Fig. \ref{fig:ST_100}
\footnote{
 The author thanks G. C. Cho for drawing this figure
several times.
} shows
a constraint implied from $S$ and $T$ parameters
\footnote{
 Both axes are essentially $S$, $T$ themselves here.
}.
 The reference standard model is at the origin on this plane
located in the contour of 90 \% confidence level.
 The slepton contribution brings $S$ parameter to negative,
while the chargino and neutralino ones to positive.
 A set of two lines in the left-hand side pursues
the response
of the slepton effects
to the change of SUSY breaking slepton mass
for two values of $\tan\beta$
(solid line for $\tan\beta = 3$,
dashed line for $\tan\beta = 50$.)
 The one on the right-hand side follows
the response of the chargino effects
against the change of $\mu$ parameter.
 The solid line with the square (triangular) marks
represents the locus followed by
the sum of the these two contributions
for the slepton mass 100 GeV (200 GeV) and $\tan\beta$ equal to 3
when $\mu$ is changed to about 500 GeV .
 Thus such a parameter set with $\tan\beta $ equal to 3
is allowed at 95 \% confidence level.
 But in the case of $\tan\beta$ = 50
it is rather difficult to take slepton mass equal
to 100 GeV.
 Once the slepton mass is taken larger,
for instance, at 200 GeV,
there is no restriction from this analysis.
\\
\quad
 Now for $\tan\beta = 3$ the contour with equal $(a_\mu)_{\rm SUSY}$
in the $\mu$-$M_2$ plane
is drawn in Fig. \ref{fig:cnt_3_100}.
 With the future accuracy, which amounts to the interval
between the neighboring contours in that figure,
we can extract SUSY effect and may obtain
precise information on the model.

\section{Discussion and Summary}
\label{sec:discussion}
\quad
 Now let us turn back to the theoretical uncertainty.
 As was mentioned by several talks in this workshop,
besides QED contribution,
$a_\mu({\rm SM})$
is also dominated by the leading order QCD contribution
which arises through the hadronic vacuum polarization.
 Its improvement is now awaiting
for the precise knowledge about the low energy
hadron production cross section
planned to be accumulated
at Novosivirsk, Frascatti and Beijing.
\\
\quad
 The hadronic light-by-light scattering effect \cite{HK},
which requires purely theoretical evaluation,
may become an obstacle.
 Thus the reduction of its error also needs
further challenge.
\\
\quad
 To summarize we discussed
the effect to muon $g-2$ from two candidates of models
each of which accesses to ``hierarchy problem''.
 We found that the potential signatures are expected
from the existence of both two candidates
by future measurement of muon $g-2$
even on account of the precision measurement
at $Z$ pole.
 But this program cannot be accomplished without
improvement in measurement of the hadron production cross section
in low energy domain.
\\

{\bf Acknowledgement} \\
\vspace{0.4cm} \\
\quad
 The author thanks S. Eidelman
for hospitality at Novosivirsk.
 He also thanks to N. Sakai for occasion visiting at TIT
after this March,
and to G. C. Cho and K. Hagiwara
for the various comments on the preparatory content of talk.

\end{document}